\documentstyle[12pt]{article}     

\def\monthyear{\ifcase\month\or
  January\or February\or March\or April\or May\or June\or
  July\or August\or September\or October\or November\or
December\fi
  \space\number\year}

\def\up#1{\leavevmode \raise.16ex\hbox{#1}}

\def\slash#1{{\mathpalette\c@ncel{#1}}}

\renewcommand{\baselinestretch}{1.17}
\newcommand{\gapproxeq}{\lower
.7ex\hbox{$\;\stackrel{\textstyle >}{\sim}\;$}}
\newcommand{\lapproxeq}{\lower
.7ex\hbox{$\;\stackrel{\textstyle <}{\sim}\;$}}

\newcounter{appendice}

\def\thefiglist#1{\section*{Figure Captions\markboth
 {FIGURE CAPTIONS}{FIGURE CAPTIONS}}\list
 {Figure \arabic{enumi}.}
 {\settowidth\labelwidth{Figure #1.}\leftmargin\labelwidth
 \advance\leftmargin\labelsep
 \usecounter{enumi}}
 \def\baselinestretch{1.1}\@normalsize
 \def\newblock{\hskip .11em plus .33em minus -.07em}
 \sloppy}

\newcommand{\be}{\begin{equation}}
\newcommand{\ee}{\end{equation}}
\newcommand{\bea}{\begin{eqnarray}}
\newcommand{\eea}{\end{eqnarray}}
\newcommand{\bean}{\begin{eqnarray*}}
\newcommand{\eean}{\end{eqnarray*}}

\def\thefiglist#1{\section*{Figure Captions\markboth
{FIGURE CAPTIONS} {FIGURE CAPTIONS}}\list
{Figure \arabic{enumi}.}
{\settowidth\labelwidth{Figure #1.}\leftmargin\labelwidth
\advance\leftmargin\labelsep
\usecounter{enumi}}
\def\newblock{\hskip .11em plus .33em minus -.07em}
\sloppy}

\begin{document}
\begin{titlepage}
\begin{flushright} RAL-93-078\\
\end{flushright}
\vskip 2cm
\begin{center}
{\bf\large Issues in Light Hadron Spectroscopy\footnote{Based on an invited
talk given at the international spectroscopy conference, Hadron '93, held at
Villa
Olmo, Como, Italy, June 21-25, 1993.
}}
\vskip 2cm
 {\bf \large D Morgan\\
Rutherford Appleton Laboratory,\\
Chilton, Didcot, Oxon,
OX11 OQX, England.}\\
\end{center}
\begin{abstract}
A high priority in light spectroscopy is to seek out and characterize various
types
of non-$(Q\bar{Q}$) meson.  The large quantity of new data now appearing will
present a great opportunity.  To identify the non-$(Q\bar{Q}$) intruders one
needs to know the regular $(Q\bar{Q}$)  pattern well; whole meson families thus
become a target for close investigation.

A powerful discovery strategy is to observe the same meson in a variety of
reactions.  Because mesons appear as resonances, other dynamics can distort the
signal in a particular decay channel.  Unitarity is the master principle for
co-
ordinating various sightings of the same resonance.  Much of the new
spectroscopic information in prospect will come from inferring two-body
dynamics from three-body final states.  Conventional methods of analysis via
the
isobar model use approximations to unitarity that need validation.

Of all the meson families, the scalars should be a prime hunting ground for
non-$(Q\bar{Q}$)'s.  Even before the advent of the new results, some revisions
of the `official' classifications are urged.  In particular, it is argued that
the lightest
broad $I=0$ scalar is a very broad $f_0$ (1000).   One unfinished task is to
decide
whether $f_0$ (975) and $a_0$ (980) are alike or different; several
non-$(Q\bar{Q}$)  scalar scenarios hinge on this.  To settle this, much better
data
on $K\bar{K}$ channels is needed.
\end{abstract}

\end{titlepage}

\section{Introduction}
I survey immediate prospects for advances in light spectroscopy mid 1993($^1$).
Leaving aside many difficult conceptual problems ($^{2}$) to which we will
have to return another year, the present focus is mesons, in particular
non-$(Q\bar{Q}$) mesons ($^3$).  To pursue non-$(Q\bar{Q}$)  candidates, we
need good information on the regular $(Q\bar{Q}$)  spectrum.  This has been
increasingly available in latter years as illustrated by the Regge plot of
natural
parity non-strange mesons shown in Fig 1.  Note the substantial progress in
resolving the spectrum of $\rho, \omega$ and $\phi$ excitations of recent years
($^4$).  This illustrates the value of being able to correlate results from
alternative production processes (Sect. 3).

Establishing the existence and measuring the properties of light mesons tends
to
be a highly non-trivial undertaking.  Recall the saga that has preceded our
present understanding of the $a_1$ (1260).   In contrast to the simplicities of
heavy quark systems, raw experimental findings can be quite misleading.  For
that reason, much of this survey is devoted to methodology.  After a lightning
tour of non (-standard) $Q\bar{Q}$ systems (Sect. 2) and of alternative
production processes (Sect. 3), I dwell in some detail on various aspects of
resonance classification and the extraction of resonant signals from experiment
(Sect. 4).  The master principle is unitarity.  A problematic aspect of topical
importance concerns the analysis of three-body final states to infer two-body
dynamics.

Many of the potential complications as to how resonances manifest themselves
are exemplified in the scalar system; it is also singled out by models of meson
composition as the likely repository for all manner of non -$(Q\bar{Q}$)
entities.
Much new spectroscopic information is now forthcoming from experiments at
LEAR and elsewhere ($^5$).  Section 5 is therefore devoted to a survey of the
pre-
existing pattern into which the new information must fit.  Some re-assessments
of the `official' classifications are urged; in particular, it is argued that
the lightest
broad $I=0$ scalar is a very broad $f_0$ (1000).

A recent claim to find evidence for a narrow $f_0$ (750) is examined and found
to be unconvincing.  Many scenarios for non-$(Q\bar{Q}$)  scalars turn on the
question: are $f_0$ (975) and $a_0$ (980) alike or different?  Accurate
$K\bar{K}$
data is needed to settle this.

Following this introduction, the sections are entitled:\\
2 $-$  Types of non (-standard) $(Q\bar{Q}$) systems; 3 $-$ Alternative
production
processes;
4 $-$  Discovering and characterizing resonances; 5 $-$ Scalar mesons:  6 $-$
Conclusions
and outlook.

\section{Types of non (-standard) $(Q\bar{Q}$)  systems}

In the following, I mainly focus on states below 1.8 GeV.

According to the quark model and QCD ($^6$), most known mesons correspond
to simple, non-relativistic $Q\bar{Q}$ compounds (to be called $M_2$ states)
which group into flavour nonets distinguished by their orbital and radial
excitations.  We are seeking states that, in one way or another, depart from
this
pattern.

Table 1 lists various types of non-standard meson and possible signatures of
non-$(Q\bar{Q}$) composition that have been suggested.  Of these latter, exotic
quantum numbers are obviously decisive.  Thus settling the resonance status of
candidates with this feature, notably the $\hat{\rho}$ (1405) $P$ wave
$\pi\eta$
resonance reported by GAMS ($^7$), is of the utmost importance.  Likewise, we
need to check on the resonance, or more likely non-resonance, status of the
broad
$I=2$  threshold enhancement seen in
$\gamma\gamma\rightarrow\rho\rho$ ($^8$).  However, theory and models
suggest and experiment seems to confirm that most non-$(Q\bar{Q}$) states will
not be thus distinguished and must be sought among ordinary $J^{PC}$ families.
Here, unusual production and decay attributes may assist in identification but
the prime discovery strategy is to establish the existence of extra states
additional
to the standard $(Q\bar{Q}$) spectrum.  The spotlight of scrutiny has therefore
to
be directed on to whole families - e.g. $I=0$ scalars below 1.8 GeV - rather
than on
individual states.  Nonetheless, the types of non-$(Q\bar{Q}$) states listed in
Table 1 have their own taxonomies which should help in classifying candidates.

The two non-$(Q\bar{Q}$) species to receive most attention either contain
additional constituent quarks or constituent gluons ($^2$) (cf. Table 1).  The
first
group comprises not only molecules ($^9$), ($M_2, M_2^\prime$) but other four
and more quark configurations like $(QQ\bar{Q}\bar{Q}$)($^{10}$), including
($N\bar{N}$) bound states ($^{11}$).  Here, I concentrate on molecules, of
which
different kinds are discussed.  All emphasise $S$-states.  The first category,
which
has a specific (concealed strangeness) flavour structure $K\bar{K}, K\bar{K}
(+c.c.$) etc.($^{12}$), provides the most popular description of the seemingly
anomalous and narrow $I=0$ and 1 scalars $f_0$ (975) and $a_0$ (980).  The
molecule picture predicts a large coupling to $K\bar{K}$ and small and equal
$\gamma\gamma$ widths for $f_0$ and $a_0$.  On this view, $f_0$ and $a_0$
are highly similar structures and we need to find alternative $I=0$ and 1
candidates for the ground state $(Q\bar{Q}$) nonet.  As we shall see, the $I=1$
spectrum is experimentally much less complicated so perhaps the more
promising channel to provide decisive information.  Other models make
different predictions and we await the verdict of experiment.

T\"{o}rnqvist ($^{13}$) has proposed a second type of molecule where, as for
the
deuteron, the constituents bind by one pion exchange (OPE).  By considering
different compositions that are allowed by the (OPE) mechanism, he arrives at
the following tentative assignments:
\bea
(\omega\omega +\rho\rho)& \longleftrightarrow& AX/f_2
(1520)?\nonumber\\
(\omega\omega -\rho\rho)& \longleftrightarrow& f_0
(1590)?\nonumber\\
K^*\bar{K}^* & \longleftrightarrow & \theta /f_0
(1710)?\nonumber\\
K\bar{K}^* & \longleftrightarrow & f_1 (1540)?\nonumber
\eea
Ericson and Karl ($^{14}$) have suggested that T\"{o}rnqvist's criterion for
binding
needs refinement.

The other group of non-$(Q\bar{Q}$) species to be considered features mesons
built wholly or partly from glue (called Chromocules in Table 1).
Both glueballs $(GG)$ and hybrids like $(Q\bar{Q} G$) arise within theoretical
schemes that describe large distance, confining QCD ($^{15}$).  Hybrids are
generally expected to be heavier than 1.8 GeV and I shall ignore them.

In order of ascending mass, the lightest glueballs should be the scalar, tensor
and
pseudo-scalar.  Rather specific predictions emerge from pure SU(3) gauge theory
calculations (i.e. omitting dynamical quarks) on the lattice; typical modern
findings are ($^{16}$):
\bea
m_{GG} (0^{++}) &=& 1550 \pm 50\; MeV\nonumber\\
m_{GG} (2^{++}) &=& 2270 \pm 100\; MeV\nonumber\\
m_{GG} (0^{-+})/m_{GG}(2^{++}) &\gapproxeq & 1.0
\eea
Insertion of dynamical quarks with realistic light quark masses may
considerably
modify these values.  Meanwhile, they afford a first guide to the masses and
especially mass-ratios that may be anticipated.  Optimists can readily find
candidates among the rich spectrum of $I=0$ scalars and tensors that experiment
provides.  Actually proving that such a candidate really is a glueball is hard,
although there are a number of properties, like having SU(3) symmetric decays,
that one would expect to observe ($^3$).

Also listed in Table 1 are states that do have a $Q\bar{Q}$ composition but of
a
non-standard type.  I first discuss the `novel hadrons' that arise in Gribov's
picture of the QCD vacuum ($^{17}$).  It is interesting to examine this new
scheme alongside more familiar molecular possibilities because of the novel
phenomena that it entails ($^{18}$).  According to Gribov, confinement is due
to
the formation of a $(q\bar{q}$) condensate involving the very light quarks $u$
and $d$.  Gribov's `novel hadrons' or `minions' are compact ($\sim$ 1 GeV$^{-
1}$) ($u\bar{u}\pm d\bar{d}$) scalar and pseudo-scalar excitations of this new
vacuum.  Suggested candidates for the scalar minions are $f_0$ (975) and $a_0$
(980) for which the following properties are predicted ($^{18}$):

\noindent $\bullet$ comparable coupling to $\pi\pi$ and
$K\bar{K}$($^{19}$)\\
\noindent $\bullet$ suppressed decay to `normal' hadrons ($\Gamma
(f_0\rightarrow\pi\pi)\sim \frac{1}{10}\Gamma
(a_1\rightarrow\rho\pi$))\\
\noindent $\bullet$\, $\Gamma (\gamma\gamma\rightarrow f_0)/\Gamma
(\gamma\gamma\rightarrow a_0$) = 25/9\\
(recall this last ratio is expected to be 1:1 on the ($K\bar{K}$) molecule
picture).
There are also predictions for various hard processes ($^{18}$).

All the above agencies yield extra states to the standard $(Q\bar{Q}$)
spectrum.
However, as emphasised by T\"{o}rnqvist ($^{20}$), the conventional nonet
mass and mixing patterns may also be appreciably distorted by final state
interactions (Fig 2).  This is likely to be most pronounced for very broad
states,
e.g. scalars (see Sect. 5 below) and this has been recently confirmed in
detailed
calculations by Geiger and Isgur ($^{21}$).  These latter authors find that,
compared to naive quark model estimates, the $I=0$ scalars experience
considerable distortions; the initially non-strange state ($u\bar{u}
+d\bar{d}$)
has its mass depressed by several hundred MeV and its initially ($s\bar{s}$)
counterpart by some 50 MeV with associated change of flavour composition to
an approximately octet make-up.  Such possibilities need to be borne in mind in
attempting to classify the $0^{++}$ spectrum delivered by experiment.  Whilst
not adding to the total number of states, it complicates the quest for
non-($Q\bar{Q}$) states by distorting the standard mass and flavour patterns.

Yet another potential source of confusion occurs where opening inelastic
channels provide a source of non-resonant enhancement.  The broad peaking
observed in $\gamma\gamma\rightarrow \rho^0\rho^0$ and $\rho^+\rho^-$
cross-sections above threshold ($^{8}$) (which if resonant entail $I=2$ as well
as
$I=0$ states) are probably of this type.

A favourite way to identify non-($Q\bar{Q}$) candidates is to discover states
that
appear to be extra to the standard ($Q\bar{Q}$) spectrum.  One therefore needs
a
comprehensive model of the `normal' ($Q\bar{Q}$) spectrum (complete with
radial excitations) to serve as a template against which to measure
abnormalities.
One such description (for other possibilities see ($^{22}$)) is provided by the
non-
relativistic potential model
of Godfrey and Isgur ($^6$) and the resulting comparison for $I=0$ scalars and
tensors is shown in Fig. 3.  The format is adapted from the excellent review of
Burnett and
Sharpe ($^3$) with the experimental information updated (details of Fig. 3 are
discussed below).  These authors show similar diagrams for $I=0$ unnatural
parity levels $1^{++}$ and $0^{-+}$, not shown here since the phenomenological
situation is essentially unchanged.  Each of the $1^{++}$ and $0^{-+}$ families
appears to possess a `spare' $I=0$ state.  The most likely non-($Q\bar{Q}$)
candidates that result are, for $1^{++}$, $f_1$ (1420), and for $0^{-+}, \eta$
(1420),
the lighter of the two states into which $\iota/\eta$ (1440) seems to be
resolved
($^{23}$).  Confirming and refining our classification of these unnatural
levels is
just as important as the corresponding exercise for the natural families but
not a
prime concern for this year from lack of new experimental input.

The low mass channels like $\eta\pi\pi$ that dominate decays of these  low
spin
unnatural parity families are quite distinct from those that couple to the
natural
parity levels of Fig. 3.  For this reason, study of the unnatural levels $-$
the
$\eta$'s, $f_1$'s and $h$'s $-$ tends to be largely decoupled from that of the
corresponding scalars and tensors.  It is on these latter that copious
information
is presently emerging from LEAR($^5$) and elsewhere  and which will dominate
the following discussion.

Fig. 3 shows the experimental situation for $I=0$ scalars and tensors prior to
this
meeting.  The reason for displaying both spectra together is that scalars and
tensors usually couple to the same final states and have to be distinguished by
somewhat fallible amplitude analysis; ambiguities and changes of $J^P$
assignment are not infrequent as we shall see.

In all but a few cases, the states shown have been accorded `confirmed' states
in
the 1992 Particle Data Tables ($^{24}$) (henceforth PDG92).  (The case for
$f_0$
(1000) is presented in Sect. 5.)   Of other
`unconfirmed' states listed in ($^{24}$), only $f_2/AX$ (1520), $f_0$ (1525)
and
$f_2$ (1810)
are included here.
`Confirmed' states are indicated either by open diamonds, $\Diamond$, or
circles, $\bigcirc$, according to whether they are conventionally viewed as
($Q\bar{Q}$) or non-($Q\bar{Q}$) candidates.  `Unconfirmed' states have
question
marks; other annotation is explained below and in the caption.  $\theta$ (1710)
(`confirmed') is one of the states for which the favoured $J^P$ assignment has
fluctuated.  PDG92 follows the MK III collaboration ($^{25}$) in revising the
original $2^{++}$ finding to $0^{++} -$ hence $f_0$ (1710); however, WA76 still
report a preference for $2^{++}$ on the basis of larger statistics ($^{26}$).
Fig. 3
shows both alternatives.  As we shall hear ($^{27}$), a similar tensor-scalar
ambiguity seems to arise for the $f_2/AX$ (1520) with a large part of the
former
pure-tensor signal re-assigned to $0^{++}$.  The $f_0$ (1525) decaying to
$K\bar{K}$ reported by LASS ($^{28}$) could be another facet of this state but
would then cease to be a natural candidate for the $(s\bar{s}$) quark model
state.
Yet another scalar signal in this same mass region is the $f_0$ (1590) of GAMS
($^{29}$).  Much more work is needed to see if $f_0$ (1525) is really distinct
from
this.

There are two other aspects of the low-mass $f_0$ spectrum
to which I return in Sect. 5.  Firstly, I re-state and amplify the
suggestion ($^{30}$) that the lightest broad $I=0$ scalar is not $f_0$ (1400)
as
recommended by PDG92 but a very broad $f_0$ (1000).  Secondly, I examine and
argue against Svec et al's claim to identify a narrow $f_0$ (750) signal
($^{31}$).

\section{Alternative Production Processes}

Getting data on mesons means studying meson resonances.  A good way to
enlarge our knowledge is to study the same final state in different production
reactions.  This is how our knowledge of vector mesons has been enlarged and
refined by collating formation experiments in $e^+e^-$ annihilation and
diffractive photo-production ($^4$).

Fig. 4 illustrates some of the major reactions that have powered meson
spectroscopy but is obviously not exhaustive.  Sundry decay processes like
$K_{e_4}$ have also provided vital information.  Different reactions have been
emphasised at different epochs as experimental facilities have evolved.
Studies
of non-diffractive peripheral reactions like $\pi N\rightarrow\pi\pi N$ thus
preceded the corresponding studies of central production.  In fact, they
provide
powerfully complementary information; this can be yet further re-inforced by
suitable data on decays.  Thus properties of $f_0 (975)$ are extracted from
joint
analysis of peripheral and central production and $D_s$ and
$J/\psi\rightarrow\phi\pi\pi (K\bar{K})$ decays ($^{30}$).  The guiding
principle for such analyses is the enforcement of unitarity (Sect. 4).

Of the various production processes illustrated in Fig. 4, some have been
singled
out for their potential selectivity of different kinds of meson.  For example,
two-
photon formation of a resonance should be directly related to its charged
constituents; the resulting relations among 2$\gamma$ widths of members of
the tensor nonet are well-fulfilled ($^{32}$).  Within the quark model, the
widths
for corresponding members of different nonets belonging to the same L-band,
e.g.
$0^{++}$ and $2^{++}$, are simply related.  A purely non-relativistic
calculation,
($^{33}$) yields a ratio $\Gamma (0^{++})_{Q\bar{Q}}/ \Gamma
(2^{++})_{Q\bar{Q}}$ = 15/4 (times relative phase space factors); relativistic
corrections are estimated to reduce the ratio to near  two ($^{34}$).  A good
way to
establish the credentials of $(Q\bar{Q})$ scalar candidates is therefore to
observe
the expected production in two-photon processes ($^{35}$) (scalar glueballs and
molecules are expected to have much smaller 2$\gamma$-widths than the
corresponding $(Q\bar{Q}$)'s($^{36}$)).  Such processes can be a discovery tool
in
their own right as exemplified by Crystal Ball's claim to see a new resonant
signal, $\eta_2$ (1870) $\rightarrow \eta\pi\pi \; 0^+(2^{-+}$), in
$\gamma\gamma\rightarrow\eta\pi\pi$ ($^{37}$).

In the same simple-minded spirit, other types of reaction should be glue-rich
and
favour the production of glueballs.  The favourite and best studied example has
been $J/\psi$ EM decay where the partonic evolution leading to the final state
meson should pass through a two-gluon intermediate state.  Although new
resonances with serious claim to be considered as glueball candidates, $\eta$
(1440) and $\theta$ (1710), were discovered in this reaction ($^{38}$), many
familiar $(Q\bar{Q}$) systems also feature (Fig. 5).  This led Chanowitz to
seek a
more discriminating criterion in the  concept of `stickiness' ($^{39}$)
$$
S = \frac{\Gamma (J/\psi\rightarrow\gamma
X)}{PS(J/\psi\rightarrow\gamma X)} \times
\frac{PS (\gamma\gamma\rightarrow
X)}{\Gamma(\gamma\gamma\rightarrow X)} ,
$$
which expresses the `two-gluon' relative to two-photon coupling.

A major source of new spectroscopic information at the present time is from
the study of $p\bar{p}$ annihilation, typically to final states comprising
three
pseudo-scalars like $3\pi^0, \pi^0\pi^0\eta$ and $\pi^0\eta\eta$ ($^{40}$).
Most of the new data are on annihilation at rest and huge statistics are
involved.
Spectroscopic information is sought from study of the pair-wise dynamics of the
final state particles, e.g. $\pi^0\pi^0,\pi^0\eta$ and $\eta\eta$, using the
isobar
model.  This imports extra uncertainties (see below), as does the fact that
both $S$
and $P$ wave $(p\bar{p}$) atomic states can usually contribute ($^{41}$).  The
rich potential of the new data makes it imperative to explore and calibrate
these
complications.

\section{Discovering and characterizing resonances}

\subsection{General Remarks}

Getting data on mesons means studying meson resonances; all but the lightest
meson decay via strong interactions.  We have to study them via their decay
fragments as we do $Z^0$ and $W^\pm$ and hope to do for the Higgs.  The need
to identify mesons as resonances in final state interactions gets harder as the
resonance widths get larger.  Resonance features get more and more entangled
with threshold effects and other `background' dynamics.  This problem has
maximum scope among scalar mesons, a family of prime interest in the quest for
non$-(Q\bar{Q}$)'s of various binds.  Once background and threshold effects
enter, the same resonance can present a markedly different appearance in
different processes.  There is an obvious risk of counting different
manifestations
as different resonances (the morning star $\neq$  evening star fallacy).  We
need
a
universal parameterization to cut through such ambiguities.  $S$-matrix
principles, especially unitarity, provide the answer and require that we should
characterize resonances by the associated poles in the complex energy plane.
Resonance poles are universal; by unitarity they occur at the same place in all
reactions to which a given resonance couples.  They yield a stable
parameterization and are therefore suitable for compilation.  This is well
exemplified by the satisfactory consistency of alternative determinations of
the
$f_0(975$) (Sheet II) pole position from  a large variety of reactions
($^{42}$) in
some of which the $f_0$ appears as a dip, in others as a peak ($^{40})$.  How a
given resonance appears in a particular process depends on background phases
and flux and phase-space factors.  All this is encapsulated in the slogan $-$
{\bf
not
all bumps are resonances and not all resonances are bumps} (cf. Fig. 6).

Now for some assorted remarks:\\
(i)  $K$-matrix poles are {\bf not} suitable objects to identify with
resonances.
\\(ii)  When several channels couple, other parameters besides the complex
resonance pole (branching ratios or equivalently coupling constants) are needed
to specify a resonance.\\ (iii)  There is no theoretical reason to disallow
very
broad resonances - quite the contrary, since such objects obviously dominate
the
corresponding cross-sections in the sense of duality.  This is not to say that
experimental claims do not need careful scrutiny.  Very broad (and for that
matter very narrow) resonances are obviously hard to detect and/or establish
($^{43}$).

\subsection{Resonances close to a 2-body $S$-wave inelastic threshold}

Resonances, especially $S$-wave resonances like $f_0$ (975) and $a_0$ (980),
that
occur close and couple strongly to an opening inelastic threshold need special
treatment ($^{30}$).  There will in general be twin poles corresponding to a
given
resonance distinguished by the `sheet-structure' of the complex energy plane
induced by the inelasticity (Fig. 7).  One pole is `below threshold'
(technically on
Sheet II in the standard convention); the other is `above threshold' (Sheet
III).
Each pole introduces a distinct mass and width
$$
E^N_R \equiv M^N_R - i\Gamma^N_R/2 \quad\quad (N = II,III)
$$
Only  in the limit of vanishing inelasticity do these poles correspond to the
same
complex energy.  Generally,   $\Gamma^{III}_R$ is greater than
$\Gamma^{II}_R$ and $E^{III}_R$ in consequence more difficult to establish and
measure.  Very accurate information is needed on the inelastic reaction;
otherwise $\Gamma^{III}_R$ is poorly determined.  If a Breit-Wigner
description applies, the associated elastic width, $\Gamma^{BW}_R$, a quantity
that is often cited,  for example in attempts to classify groups of particles
with the
same $J^P$, is approximately related to the above $\Gamma^N_R$ by the
formula ($^{30}$)
$$
\Gamma^{BW}_R = 1/2 (\Gamma^{II}_R + \Gamma^{III}_R)
$$
The difficulty of fixing $\Gamma^{III}_R$ explains why widths ranging from
sixty to several hundred MeV are ascribed to $a_0 (980$) on the basis of the
same data.
  Analysis of this
resonance, and of its companion $f_0(975)$, is seriously handicapped for want
of
accurate measurements of its $K\bar{K}$ decay.
As an example of the problems that arise when $K\bar{K}$ information is
lacking, the $a_0(980)\rightarrow\pi\eta$ signal seen in
$p\bar{p}\rightarrow\pi\eta\eta$  ($^{44}$) is very well fitted assuming {\bf
zero} coupling to $K\bar{K}$.

For resonances that occur just below an inelastic threshold, counting nearby
poles can distinguish molecules from regular quark model states and
chromocules ($^{45,46}$).  A molecular resonance generated dynamically by the
scattering forces between the individual `atoms' (e.g. $K$ and $\bar{K}$) can
have just one nearby pole; the existence of two nearby poles on sheets II and
III
 points to some other dynamical origin.  Thus, from a recent analysis of a
wide variety of processes coupling to $f_0(975)$, Michael Pennington and I
concluded that present data disfavour a molecular interpretation for this state
($^{46}$).

\subsection{Analysis of multi-channel, multi-reaction data enforcing $S$-matrix
constraints}

This is a field in urgent need of fresh ideas.  The basic principles -
unitarity,
analyticity and the like - are not in doubt; the problem is how to implement
them adequately yet practicably in the situations that we actually encounter.
Reactions that are confined to at most a pair of two-body channels give no
difficulty.  Problems enter when we have to allow for three, four or more
channels and also where three (and more) body final states occur.  Both of
these
complications are present in many of the situations that dominate contemporary
spectroscopy.  For example, a key issue in scalar spectroscopy (Sect. 5) is to
decide
how many distinct $I=0$ resonances there are in the region 1.0 to 1.8 GeV.
Many
signals have been reported in a variety of reactions yet, even at 1.6 GeV, the
number of (effective) channels is at least five.  How, using unitarity, is all
this
information to be correlated?  The second complication is equally pressing: how
three-body final states can be a reliable source of information on two-body
dynamics is central to the exploitation of the wealth of data now available on
$p\bar{p}$ annihilation ($^{40}$).

Recall the standard procedure when the final states are few and simple.  To
establish the existence and properties of a resonance, $R$, we would ideally
like
to know the partial wave scattering amplitude $T_{ij}$ connecting all the
channels that couple to $R$.  Given a complete set of data on all the relevant
$\sigma_{ij}$ (along with suitable phase information from interference with
other partial waves), we would fit to a unitarity enforcing parameterization
such
as that provided by the $K$-matrix ($^{47}$)
$$
\mathop{T}_{\sim} = \mathop{K}_{\sim} (\mathop{1}_{\sim} -
i\mathop{\rho}_{\sim} \mathop{K}_{\sim})^{-1}
$$
Once the number of channels with significant coupling exceeds two, a general
$K$-matrix parameterization ceases to be practical and alternative ways of
enforcing the major consequences of unitarity have to be found.  Some authors
have represented $\mathop{K}\limits_{\sim}$ by a sum of pole contributions
which are separately unitarized with neglect of `cross-talk' between the
resulting
resonant terms ($^{48}$).  Once resonances are broad and overlapping, such a
representation is almost certainly inadequate.

In practice, even when there are just a few channels, data on the $T_{ij}$'s is
always insufficient and can be usefully supplemented by information on
associated {\bf production processes}.  Where these are {\bf non-strongly
interacting}, unitarity (indicated schematically in Fig. 8) shows that we can
express the associated production amplitudes, $F_i^{(p)}$, in terms of the
$T_{ij}$ via the relations ($^{49}$)
$$
F_i^{(p)} (E) = \sum_j \alpha_j (E) T_{ij} (E)
$$
where {\bf the $\alpha_j (E)$ are smooth real functions of energy}.  This form
guarantees that resonance poles feed through to the $F_i^{(p)}$ and enables the
often very precise and fine-grained information from such production processes
to be fully harnessed.

Applications usually involve some retreat from the `non-strongly interacting'
requirement for the production processes used.  A common situation is where
there are additional final state particles to those whose dynamics is studied;
for
example, information on $\pi\pi$ and $K\bar{K}$ dynamics is extracted from
the reactions $J/\psi\rightarrow\phi\pi\pi (K\bar{K})$ treating the $\phi$ as a
spectator ($^{30}$).  In most such cases, the effect of this approximation is
likely to
be small but
important questions do arise in one key application $-$ the analysis of three
body
final states via the isobar model.

This has crucial relevance to the extraction of spectroscopic information from
$p\bar{p}$ annihilation at rest to three body final states like $3\pi^0,
\pi^0\pi^0\eta$ and $\pi^0\eta\eta$ ($^{40}$).  The dynamics studied is that of
the various {\bf pairs} of final state particles, $\pi^0\pi^0, \pi^0\eta$
and $\eta\eta$.
Analysis is based on the isobar model whereby the three-body production
amplitude (for each atomic partial wave) is firstly written as a sum of three
terms
(see lower portion of Fig. 4)
$$
F^{123}=
F^{1;23} + F^{2;31} + F^{3;12}
$$
classified according to which pair interacted last.  Each term is a sum of
partial
wave amplitudes $F_{L_{23}}^{1;23} (s_{23})$, that are subject to 2 and 3 body
unitarity requirements.  The isobar model assumes that `crossed re-scattering
effects' from `triangle diagrams', where one of the emerging pair constituents
re-
groups with the associated spectator (1(23)$\rightarrow$ 123 $\rightarrow$
(12)3), are unimportant.  Each isobar component then conforms to the previously
considered case with two interacting final state particles and a spectator
allowing
one to write (omitting the angular momentum label $L_{23}$ and restoring the
previous channel labels $i$ and $j$)
$$
F^{1;23}_i(s_{23})= \sum_j \alpha_j (s_{23}) T_{ij} (s_{23})
$$
with the pre-factors $\alpha_j$ again real and slowly varying.  Given the great
spectroscopic potential of the $p\bar{p}$ data now being analysed, crossed re-
scattering corrections to the isobar approximation should be evaluated, at
least
for selected examples, using standard methods ($^{50}$).

\section{Scalar Mesons}

As we have seen, almost all mechanisms for generating meson resonances
predict light scalars; in some cases, only scalars are expected in the low mass
region.  For this reason alone they are of exceptional interest.  Add to this
the
large mass of new data coming on stream and one sees why scalars are this
year's
most exciting topic.  This last section is therefore devoted to some clearing
of the
ground in preparation for the new results.

I begin with a quick survey of the `official' $0^{++}$ spectrum according to
PDG92
($^{24}$).  I then focus on two particular questions relating to the $I=0$
spectrum:
firstly, I examine and argue against Svec et al's ($^{31}$) claim to identify a
narrow $f_0$ (750) signal in peripheral dipion production; then, I restate and
amplify the assertion ($^{30}$) that the lightest broad $I=0$ scalar is not
$f_0$
(1400) as recommended by PDG92 but a very broad $f_0$ (1000).  This has
important consequences for our perception of where we believe the
$(Q\bar{Q}$) scalars
cluster in mass.  I end with some general questions and comments.

According to PDG92, the spectrum of scalars below 1800 MeV comprises the states
shown in Fig. 9 $-$ excepting of course the $f_0$ (1000).  (Notation as for
Fig. 3 in
which $I=0$ scalars were already displayed $-$ cf. related discussion in Sect.
2)
How does this spectrum accord with what we might or should expect?  Given the
success of the naive quark model description of the other nonets, 2$^{++},
1^{++}$ and $1^{+-}$, of the $L=1$ band, we should certainly expect to find an
analogous $0^{++}(Q\bar{Q})$ nonet.  We need to equip this with the standard
$I=1/2 (K_0), I=1 (a_0)$ and pair of $I=0(f_0)$  members (the latter may or may
not be ideally mixed).  Over the years, opinion has fluctuated as to which of
the
available states provide the most likely occupants for these slots.  In all
these
gyrations, the $K_0$ (1430) has been a fixture (its mass used to be somewhat
lower $-$ indeed the LASS group ($^{51}$) who are the source of the present
Table
values, report a second fit yielding a mass of 1350 MeV, as indicated in Fig.
9).  At
first, the known scalars were just sufficient to populate a nonet using the
broad
$f_0(\varepsilon), K_0, f_0$ (975) and $a_0$ (980)($^{52}$).  Later the
prevailing
opinion
came to be that these last two were too light and too narrow to be plausible
$(Q\bar{Q})$ candidates (despite arguments that final state interactions can
induce exceptional mass and mixing shifts for the scalars($^{20,21}$)and the
ambiguities in the concept of resonance width for such near threshold states
(cf.
Sect. 4.2)).  Then came the suggestion that $f_0$ (975) and $a_0$ (980) could
be
$K\bar{K}$ molecules ($^{12}$).
Finally possible substitute candidates for the vacated $(Q\bar{Q})$ slots were
reported in the guise of $a_0$ (1320), inferred from analysis of $\pi\eta$
production ($^{53}$) and $f_0$ (1525) seen in $K\bar{K}$ ($^{28}$).  It was
pointed
out that, given these replacements, the ensuing $(Q\bar{Q})$ scalar nonet would
closely resemble its other $L=1$ companions - an attractively simple synthesis
($^{54}$).  The empirical evidence for the new states is far from compelling.
Each
relies on amplitude analysis of a single experiment leading to a scalar signal
with
the same mass and width as a co-present and dominant tensor state.  Whether or
not these two signals are confirmed, the existence of alternative $I=0$ and 1
scalar $Q\bar{Q}$'s is of great importance for our understanding of the quark
model $-$ hence the interest in new $a_0$ and $f_0$ signals now being reported
($^{27,40}$) (the new $a_0$ signal  is indicated in Fig. 9).

The two remaining states shown in Fig. 9, $f_0$ (1590) and $f_0$ (1710), both
raise
very interesting questions to which I return after discussing $f_0$
spectroscopy at
lower energies.

\subsection{Resonances seen in $\pi\pi$ and $K\pi$ phase shifts}

As we have seen, properties of the narrow scalars, $f_0 (975)$ and $a_0(980$),
can be  investigated in a whole variety of reactions in which they appear.  For
$I=1/2$ and other $I=0$ dynamics, we must mainly rely on phase shift analyses
based on peripheral di-meson production assuming OPE dominance ($^{55}$).
(The lack of a similar direct window on $\pi\eta\; I$=1 phase shifts may
perhaps
be remedied by a careful study of $\pi\pi\eta$ final states ($^{56}$).)

Fig 10(a) shows the well-known form of the $K\pi\; S$-wave phase shift
($^{51}$) from which the $K_0$ (1350-1430) resonance is inferred with width
$$
\Gamma_{K_0} = 290-350 MeV
$$
where I have indicated the spread of values from both resonance fits reported
($^{51}$).  From this, we learn that {\bf broad $S-$wave resonances occur}, a
fact
that has implications throughout the scalar nonet.  Interpreting $K_0$
conventionally as the $(s\bar{n}$) component of an ideal nonet implies that the
corresponding $I=0$ ($u\bar{u}+d\bar{d}$) state decays to $\pi\pi$ with
approximately double the above width (and also reinforces doubts concerning
the reported $a_0$ (1320) $\rightarrow \pi\eta$ and $f_0$ (1525) $\rightarrow
K\bar{K}$ signals as being too narrow).  So what do we learn from the
corresponding $I=0$ phase shifts?

The accepted form of this phase shift, $\delta^0_0$, from threshold to 1.4 GeV
($^{24}$) is plotted (modulo 180$^0$) as the full line in Fig. 10(c).  Before
going
into the interpretation of this, I briefly examine the challenge by Svec et al
($^{31}$)  to this description of $\delta^0_0$ for dipion masses up to 900 MeV.
 It
is first necessary to supply some historical background.

Prior to the high statistics dipion ($\pi^+\pi^-$) production experiments of
the
early 70's, discussion of the $I=0\;\pi\pi\;S$-wave phase shift below 1 GeV was
beset by an UP-DOWN ambiguity ($^{57}$)  (Fig. 10 (b) ($^{58}$)).  This arose
because the $S-$wave in $\pi^+\pi^-$ production is inferred from interference
with the dominant $(P$-wave) $\rho$ signal.  In principle, this should have
been resolved by study of $\pi^0\pi^0$ production but the corresponding
experiments have delivered a conflicting verdict ($^{59}$).  By common consent,
the matter was resolved with the DOWN alternative selected once the $f_0(975$)
signal was clearly de-lineated ($^{60}$) .  Svec et al. ($^{31}$)  claim to
resuscitate
the UP solution in an amplitude analysis of their own and earlier Cern-Munich
(CM)  ($^{61}$) dipion  production experiments off polarised targets for
$\pi^+\pi^-$ in the mass range 600-900 MeV (That this analysis stops at 900 MeV
is a significant limitation.)  From the CM data at small $t$,
they find an UP and DOWN solution.  Their own data at larger $t$ only yields
evidence for the UP alternative.  From this, they infer the existence of
$$
f_0 (750) \qquad \Gamma (100-250) MeV
$$
citing Cason et al's 1983 $\pi^0\pi^0$ results ($^{59}$)  in support.

I do not think $f_0$ (750) can possibly be a real effect for the following
reasons:\\
(a)  Absence of corresponding signals in $\gamma\gamma, $ central production
and
sundry decay processes $-$ and, even more compellingly $-$\\
(b)  The requirement to join on to the $f_0(975)$ signal in  $\pi^+\pi^-$.  How
this works was spelt out by Pennington and Protopopescu back in 1973 ($^{62}$).
Using Roy's equations, they show that given the existence and observed
properties of $f_0(975)$, the DOWN solution does and the UP solution does not
reproduce itself through the associated dispersion relations.  I conclude that
we
can forget about $f_0$ (750).

So we are back to the standard form for the $I=J=0$ phase-shift shown in Fig.
10(c) and ready to address the question: {\bf what is the lightest broad $I=0$
scalar}?  For reasons that I examine below, PDG92's answer is $f_0$ (1400) with
width $\Gamma$ = 150-400 MeV.  In contrast, Michael Pennington and I
($^{30}$)  unhesitatingly plump for something in the range($^{63}$)
$$
f_0 (1000) \quad - \quad {\rm width} \; \Gamma \simeq 700 \; MeV\quad  -
$$
for the following reasons:\\
(i)  Analysis of $\pi\pi$ phase shifts from threshold to 1.4 GeV along with an
extensive set of related reactions (the AMP analysis ($^{47}$)) yields a
resonance
pole at (900-$i$350) MeV (c f. Hyams et al's 1973 result of (1049 -$i$250) MeV
($^{64}$)).\\
(ii)  For an intuitive feel as to what is going on($^{30}$),  take the standard
form
of $\delta^0_0$, as shown in Fig. 10(c) (full line) and `remove' the rapid
$f_0(975)$ phase excursion.  One then sees a slow, steady ascent of the
residual
phase shift (dashed line) just like $\delta_{K\pi}$ (Fig. 10(a)).\\
(iii)  Our resonance spectrum now accords well with the weighted mean of the
partial-wave cross-section in line with notions of duality.

How does this square with PDG's $f_0$ (1400) and their casting it in the role
of
lightest broad $f_0$?  PDG base their recommendation ($^{24}$)  on an
assortment of resonance signals derived from $\pi\pi,K\bar{K}$ and $\eta\eta$
final states ($^{65,66,29}$).  They appear to place most reliance on the paper
reporting the AFS experiment on $pp\rightarrow pp\pi^+\pi^-$ ($^{65}$); in
particular, they cite an amplitude analysis therein that `shows that [the]
$\pi\pi\;S$-wave dominates up to 1.6 GeV with no room left for other scalars
besides $f_0$ (975) and $f_0$ (1400)!'  The first thing to say is that the
analysis in
question is confined to the subset of data with $M_{\pi\pi}$ above 1 GeV (i.e.
after the first precipitate fall of the $\pi\pi$ spectrum).  The whole $\pi\pi$
spectrum from threshold to 1.4 GeV  has in fact been well-fitted along with a
large quantity of other $I=0$
data in the AMP analysis referred to above ($^{47}$).  Far from excluding $f_0$
(1000), it strongly reinforces it.  What the AFS analysis does do is indicate a
second $f_0$-resonance signal at $(M=1420\pm$ 20 MeV with width $\Gamma
=460\pm$ 50 MeV), seemingly from the need to fit the second dip in their
$\pi\pi\;S$-wave spectrum ($^{65}$).

Other evidences for $f_0$ (1400) cited by PDG come from an assortment of
$K\bar{K}$ ($^{66}$) and $\eta\eta$ ($^{29}$) production experiments, in each
case after amplitude analysis to isolate the $S$-wave signal.  The $K\bar{K}$
experiments disagree even on the qualitative form of the $S$-wave cross-section
below 1400 MeV but mostly concur in finding narrow $f_0$ signals above 1400
(e.g. Etkin et al. ($^{66}$) report $f_0$ (1463), width 118 MeV).  Analysis of
the
$\eta\eta$ experiment ($^{29}$) yields a two hump $S$-wave from which the
authors derive a pair of $f_0$-resonances.  The upper hump provides one of the
major evidences for $f_0$ (1590) to be discussed below.  PDG suggest that the
lighter GAMS resonance - $f_0$ (1220), width 320 MeV - is another facet of
$f_0$
(1400).  Given the proximity of the lower peak to $\eta\eta$ threshold, it
would
seem more natural to make the link to $f_0$ (1000);  only multi-channel fits
can
decide.

What all this adds up to is persuasive evidence for extra $S$-wave structure
above say 1200 MeV without specifying what that structure actually is.  For
that,
we must mostly await the new (and future) data and comprehensive analyses
that include them.  However, further $f_0$ signals are already claimed $-$ not
only
$f_0$ (1525) $\rightarrow K\bar{K}$ from LASS ($^{28}$) (already discussed),
but
$f_0$ (1590) $\rightarrow\eta\eta$ (and other channels) from GAMS ($^{29}$)
and the scalar metamorphosis of the $\theta, f_0$ (1710) ($^{25}$).  Each of
these
last two could occupy key positions in our final classification.  Thus, $f_0$
(1710)
(provided its scalar/tensor spin ambiguity is final resolved in favour of
scalar)
could be the first $f_0$ radial recurrence, whilst $f_0$ (1590) has been
proposed as
a candidate for the scalar glueball.  As evidence for this latter assignment,
GAMS
($^{29}$) especially emphasise the 1590's preference for $\eta\eta$ (and
according
to them a fortiori $\eta\eta^\prime$) decay modes (however this is from the
standpoint of a particular non-standard model of scalar glueball decays).  The
more conventional expectation would entail a straight-forwardly singlet decay
pattern without the avoidance of $\pi\pi$ and $K\bar{K}$ decay modes that
GAMS stress.  It seems not unlikely that, in the final analysis, the GAMS decay
modes, $\eta\eta,\eta\eta^\prime$ and $4\pi$, will turn out after all to have
$\pi\pi$ and $K\bar{K}$ counterparts, perhaps shifted in mass; if so, it will
be
very interesting to see what $\pi\pi:K\bar{K}:\eta\eta$ ratios prevail.  What
appears beyond doubt is that there are surplus scalars $-$ thus
non$-(Q\bar{Q})$
candidates.  Only future data and careful analyses will tell us how many.

I conclude with a pair of questions that are central to how we view the scalar
spectrum overall:

(i) {\bf Is $a_0$ (980) the only $I=1\;S$-wave object below, say, 1600 MeV or
is
there something else}?  This has clear and important (but not decisive - see
(ii)
below) bearing on the identity of the ground state $I=1 (Q\bar{Q})$.  Attention
has previously been focused on the GAMS $a_0$ (1320) signal ($^{53}$) (whose
empirical shortcomings have already been described).  Now we are offered an
alternative candidate at higher mass by Crystal Barrel ($^{27,40}$).

(ii) {\bf Are the $f_0$ (975) and $a_0$ (980) alike or different}?  Very
different
scenarios would ensue if either or both of $f_0$ and $a_0$ were shown to have a
large `true' width in the sense of Sect. 4.2.  As illustration, suppose $f_0$
is
confirmed as `narrow' ($^{30,45,46}$) but $a_0$ is found to be `broad' as
several
authors have suggested ($^{67}$).  Not only would this obviously kill the
molecule and `minion' interpretations of $f_0$ and $a_0$ but, depending on the
width and branching ratios actually found, could allow $a_0$ (980) to be
reconsidered as a candidate for the $I=1 (Q\bar{Q})$ ($^{68}$).  At present, we
cannot rule out such a possibility because we do not really know
$\Gamma_{BW}(a_0) -$ for lack of accurate data on $a_0\rightarrow
(I=1)K\bar{K}$.  Likewise we need better information on $(I=0)K\bar{K}$ to
check conclusions on $\Gamma_{BW}(f_0)$.  This highlights the pervasive
need for improved $K\bar{K}$ data.

\section{Conclusions and outlook}

That concludes my pre-Como tour of the light meson.  What lessons emerge?

First, whilst stressing the key role of unitarity in parameterizing and co-
ordinating various resonance signals, I noted the limitations of present
practice.
Once the number of channels grows (essentially beyond two), or three-body final
states, except of very restricted type, enter, present methods are either
inadequate
or impractical or both.  Here is one area calling for fresh ideas.

I touched on the great variety of production  process that can bear on meson
spectroscopy.  Some appear to offer exceptional promise for future
exploitation.
Two photon production could provide a powerful means of probing $C=+$
mesons and have great potential for discriminating alternative compositions.
Present data, being a by-product of $e^+e^-$ annihilation studies, is limited
in
scope.  Custom built photon-photon facilities ($^{69}$) could transform this.
Another promising and expanding area is central production, with its ability to
produce well-isolated samples of a whole variety of meson final states, not
only
of natural but also of unnatural parity ($^{70}$).  Production systematics need
much more study in order to exploit this resource to the full.

A key area for this year is the family of scalars.  Surveying the pre-existing
information, I restated the argument ($^{30}$) that the lightest $I=0$ scalar
is a
very broad $f_0$ (1000) and stressed the need for more information on the
ubiquitous $f_0$ (975) and $a_0$ (980) systems especially in their $K\bar{K}$
final states.  Once the new results ($^{5,27,40}$) are assimilated, we will
need to
take stock of the enlarged scalar spectrum that emerges.

For contingent reasons, the emphasis this year has been on natural parity
states,
however possible $J^P=0^-$ and $1^+$ non $(Q\bar{Q}$) candidates like $\eta$
(1420) and $f_1$ (1420) are just as interesting and also need much more
investigation.  Light spectroscopy is a seamless web and we need  advance on
 all fronts to grasp the overall design.

\section*{Acknowledgements}

It is a pleasure to thank: Michael Pennington for a long standing collaboration
in
which much of the distinctive outlook presented here was developed; Iain
Aitchison, George Gounaris, Wolfgang Ochs, Antimo Palano and Michael Teper
for very helpful communications; various members of the Crystal Barrel
collaboration for enlivening discussions; and, finally, the organisers of
Hadron
'93 for arranging a very stimulating meeting.

\newpage
\subsection*{Table 1:   Types of non(-standard) ($Q\bar{Q}$) configuration}

In addition to the regular $M_2 \equiv$ ($Q\bar{Q}$) mesons of the non-
relativistic quark model which group into (mostly ideal) flavour nonets
distinguished by their orbital and radial excitations, we may have:

\noindent $\bullet$ \quad {\bf MOLECULES} ($M_2, M_2^\prime$) $-$ and
other four and more quark configurations

\noindent $\bullet$\quad  {\bf CHROMOCULES} $-$ glueballs $(GG)$, hybrids
($Q\bar{Q}G$)  etc.

If particular mechanisms operate, non-standard types of ($Q\bar{Q}$)  system
can
arise:

\noindent $\bullet$ \quad {\bf GRIBOV'S}  `novel hadrons' (OR {\bf
`MINIONS'})
($>Q\bar{Q}<_0$)

\noindent $\bullet$ \quad {\bf HEAVILY RENORMALIZED }($Q\bar{Q}$)'s

(These can occur where resonances have very large widths leading to nonet mass
and flavour patterns being appreciably distorted by final state interactions.)

\subsection*{Non-($Q\bar{Q}$) Signatures}

\noindent $\bullet$ \quad {\bf EXOTIC QUANTUM NUMBERS}

\noindent $\bullet$ \quad {\bf UNUSUAL PRODUCTION AND DECAY
PROPERTIES}

\noindent $\bullet$ \quad {\bf SPARE STATES}

Potentially misleading signals for ($Q\bar{Q}$) or non ($Q\bar{Q}$) states
could
come from:

\noindent $\bullet$ \quad {\bf NON-RESONANT ENHANCEMENTS FROM
OPENING CHANNELS.}
\newpage

\begin{thefiglist}{25}
\item{ } Regge plots for natural parity non-strange mesons listed in ref
($^{24}$).
\item{ } Primordial quark model states may be modified by final state
interactions ($^{20}$).
\item{ } $I=0$ scalar and tensor mesons.  Observed states (pre-Como) compared
to quark model predictions ($^6$) - see text for details.  The case for
replacing
$f_0$ (1400) by $f_0$ (1000) as the lightest broad $I=0$ scalar is given in
Sect. 5.
\item{ } Alternative production processes.
\item{ } Branching ratios for $J/\psi\rightarrow\gamma M \times 10^3$ for
various mesons $M$ versus mass (numbers taken from ref. ($^{24}$), $\ddag$
the $\eta$ (1440) entry is via the $K\bar{K}\pi$ mode and $f_0$ (1710) via
$K\bar{K}$).
\item{ }Example of how bumps need not correspond to resonances nor
resonance signals appear as bumps $-$ AFS data on $\pi^+\pi^-$ central
production ($^{65}$) and the corresponding AMP analysis fit ($^{47}$).
\item{ } Resonance and bound state poles and the sheet structure of the energy
plane $-$ how their relation is clarified by mapping onto a suitable $k$-
(momentum) plane: (a) and (b) depict one channel examples as for the deuteron
(bound state (a)) or corresponding spin-singlet (anti-bound state (b)); (c) and
(d)
show two-channel examples with $k_2$ the CM momentum of the inelastic-
channel (two-body channels assumed throughout).  For resonances like $\rho$
(770) or $f_2$ (1270) far from inelastic threshold, the identity of the
physically
relevant nearby pole is unambiguous  (c); for $S$-wave resonances close to
inelastic
threshold, the sheet structure ((d) and (e)) matters  (d).  Such resonances in
general have `below threshold' (Sheet II) and `above threshold' (Sheet III)
poles.
Molecular resonances arising from intra-hadron forces of finite range
only have one nearby pole, as happens for the deuteron (a).
\item{ } Diagrams to illustrate unitarity constraints $-$ (a) among a set of
scattering
amplitudes ($T_{ij}$) describing strong transitions between a set of
connecting channels $(i,j =1\ldots n$); (b) how amplitudes $F^{(p)}_i$
describing
non-strongly interacting production processes $(p)$ leading to the same set of
channels $(i$ = 1 $\ldots n$)
 are related to the $T_{ij} -$ for details see refs. ($^{47,30}$).
\item{ } Scalar meson spectrum pre-Como (for update see refs. ($^{5,27,40}$)).
States shown are those listed in PDG'92 ($^{24}$) plus the very broad $f_0$
(1000)
($^{30}$) argued for in the text.  PDG's `confirmed' states are indicated
either by
open diamonds, $\Diamond$ or circles, $\bigcirc$, according as they are
conventionally viewed as $(Q\bar{Q}$) or non$-(Q\bar{Q}$) candidates.  States
that are `unconfirmed' or whose spin is controversial have question marks.
Dashed lines indicate possible shifts of assignment mentioned in the text: (a)
alternative parametrization of the $K_0$; (b) substitution of the newly
reported
$a_0$ signal ($^{27,40}$) for the `unconfirmed' $a_0$ (1320) ($^{53}$); (c)
replacement of $f_0$ (1400) by $f_0$ (1000) as candidate for the lightest $I=0
(Q\bar{Q}$).
\item{ } Aspects of $S$-wave $K\pi$ and $\pi\pi$ phase-shifts:\\ (a)
$\delta_{K\pi}(I=1/2)$ according to the LASS experiment ($^{51}$); (b) the old
UP-
DOWN ambiguity of $\delta^0_0$($^{58}$); (c) the accepted modern form of
$\delta^0_0(I=0)$  from threshold to 1.4 GeV ($^{24}$) plotted
modulo 180$^0$ (full-line), and the residual phase after removal of the $f_0$
(975) signal (dashed line).  This latter corresponds to  the very broad
$f_0(1000)(^{30}$).
\end{thefiglist}


\begin{thebibliography}{78}
\bibitem{ } As concerns experimental results, the time reference of this review
is
pre-Como.  For new results - see these proceedings.
\bibitem{ }  Note the distinction between the ratio of couplings and the
observed
branching ratios.  The former comparison has relative phase space factors
removed.
\bibitem{ } Given the large width, we set the mass to 1000 MeV as a `round
number'.
\bibitem{ } $a_0$ (980) is strongly and conspicuously produced in a whole
variety
of reactions (e.g. it is the only scalar meson to produce a visible signal in
$\gamma\gamma$ reactions).  This might be taken to indicate a minimally
complicated composition $-$ ($Q\bar{Q}$) rather than (($Q\bar{Q}$)
($Q\bar{Q}$)).
\end{thebibliography}
\end{document}